\documentstyle[12pt,newlfont,amssymb,epsfig]{article}
\begin{document}
\begin{flushright}
\begin{tabular}{r}
UWThPh-1997-17\\
DFTT 44/97\\
hep-ph/9707372
\end{tabular}
\end{flushright}
\begin{center}
\Large \bfseries
SCHEMES OF NEUTRINO MIXING
FROM THE RESULTS OF
NEUTRINO OSCILLATION EXPERIMENTS
\footnote{
Talk presented by C. Giunti at the
$9^{\mathrm{th}}$ International School
\emph{Particles and Cosmology},
Kabardino Balkaria,
Baksan Valley,
Russia,
April 15--22, 1997.}
\\[0.3cm]
\large \mdseries
S.M. Bilenky$^{\mathrm{(a)}}$,
C. Giunti$^{\mathrm{(b)}}$
and
W. Grimus$^{\mathrm{(c)}}$
\\[0.3cm]
\itshape
\large
$^{\mathrm{(a)}}$
\normalsize
Joint Institute for Nuclear Research, Dubna, Russia
\\[0.3cm]
\large
$^{\mathrm{(b)}}$
\normalsize
INFN, Sezione di Torino, 
and
Dipartimento di Fisica Teorica,
\\[0pt]
Universit\`a di Torino,
Via P. Giuria 1, I--10125 Torino, Italy
\\[0.3cm]
\large
$^{\mathrm{(c)}}$
\normalsize
Institute for Theoretical Physics, University of Vienna,
\\[0pt]
Boltzmanngasse 5, A--1090 Vienna, Austria
\\[0.3cm]
\upshape
\large
Abstract
\\
\normalsize
\begin{minipage}[t]{0.9\textwidth} \small
The mixing of three and four massive neutrinos is considered.
It is shown that the neutrino oscillation data
are not compatible with a hierarchy of couplings
in the three-neutrino case.
In the case of four neutrinos,
a hierarchy of masses
is not favored by the data.
Only two schemes with two pairs of close masses
separated by a gap of the order of 1 eV
can accommodate the results of all experiments.
If the existing indications in favor of neutrino oscillations
will be confirmed,
it will mean that the general features of neutrino mixing
are quite different from those of quark mixing.
\end{minipage}
\end{center}

Forty years after its proposal by B. Pontecorvo
\cite{Pontecorvo57},
neutrino oscillations
(see Ref.\cite{BP87-CWKim})
are considered today
as one of the most interesting phenomena
in high energy physics
and one of the most promising methods
for the search of new physics beyond
the Standard Model
(see Ref.\cite{Mohapatra-Pal}).

In this report we discuss which information
on the neutrino mass spectrum and mixing parameters
can be obtained from the results of
neutrino oscillation experiments.
No evidence
of neutrino oscillations was found in many experiments
and their results are useful
in order to constrain the
allowed values of the neutrino masses and mixing parameters.
These are all short-baseline (SBL)
neutrino oscillation experiments
with reactor and accelerator neutrinos.
In particular,
we use the exclusion plots obtained from the data of the
Bugey \cite{Bugey95}
$\bar\nu_e$
disappearance experiment,
of the
CDHS and CCFR
\cite{CDHS84-CCFR84}
$
\stackrel{\makebox[0pt][l]
{$\hskip-3pt\scriptscriptstyle(-)$}}{\nu_{\mu}}
$
disappearance experiments
and of the
BNL E734,
BNL E776
and
CCFR
\cite{BNLE734-BNLE776-CCFR96}
$
\stackrel{\makebox[0pt][l]
{$\hskip-3pt\scriptscriptstyle(-)$}}{\nu_{\mu}}
\to
\stackrel{\makebox[0pt][l]
{$\hskip-3pt\scriptscriptstyle(-)$}}{\nu_{e}}
$
appearance experiments.

There are three experimental indications
in favor of neutrino oscillations.
They come from the existence of
the solar neutrino problem,
the atmospheric neutrino anomaly
and from the results of the LSND experiment.
The solar neutrino problem
is the oldest and most widely believed
indication in favor of neutrino oscillations:
the event rates measured by
all solar neutrino experiments
(Homestake,
Kamiokande,
GALLEX,
SAGE and Super-Kamiokande
\cite{solarexp})
are significantly smaller than those
predicted by the
Standard Solar Model
\cite{SSM}.
This deficit can be explained
with oscillations of solar $\nu_e$'s
into other states
and indicates a mass-squared difference\footnote{
Here the symbol
$\delta{m}^{2}$
indicates a generic difference
between the squares
of two neutrino masses.
}
$
\delta{m}^{2}
\approx
10^{-5} \, \mathrm{eV}^2
$
in the case of resonant MSW transitions
\cite{MSW}
or
$
\delta{m}^{2}
\approx
10^{-10} \, \mathrm{eV}^2
$
in the case of vacuum oscillations.

The atmospheric neutrino anomaly
has been found in the
Kamiokande,
IMB,
and
Soudan
\cite{Kamiokande-IMB-Soudan}
experiments.
It can be explained by
$
\stackrel{\makebox[0pt][l]
{$\hskip-3pt\scriptscriptstyle(-)$}}{\nu_{\mu}}
\to
\stackrel{\makebox[0pt][l]
{$\hskip-3pt\scriptscriptstyle(-)$}}{\nu_{x}}
$
oscillations ($x\neq\mu$)
with 
a mass-squared difference
$
\delta{m}^{2}
\approx
10^{-2} \, \mathrm{eV}^2
$.
However,
the existence of an atmospheric neutrino anomaly
is a controversial issue,
because no anomaly was observed 
in the ratio
of contained muon-like to electron-like events
measured in the
Fr\'ejus
and
NUSEX
\cite{Frejus-NUSEX}
experiments
and in the flux of upward-going muons
measured in the
Kamiokande,
IMB,
Baksan
and
MACRO
experiments
\cite{up-mu}.
The existence of an atmospheric neutrino anomaly
will be checked in the near future
by the Super-Kamiokande \cite{K2K}
experiment.
Moreover,
several long-baseline (LBL) experiments
will search for neutrino oscillations
due to
$
\delta{m}^{2}
\approx
10^{-2} \, \mathrm{eV}^2
$:
the
CHOOZ
and
Palo Verde
\cite{CHOOZ-PaloVerde}
LBL
$\bar\nu_e$
disappearance experiments with reactor anti-neutrinos
and the accelerator
KEK--Super-Kamiokande (K2K) \cite{K2K},
Fermilab--Soudan (MINOS) and
CERN--Gran Sasso (ICARUS)
\cite{MINOS-ICARUS}
LBL
$\nu_{\mu}$
disappearance
and
$\nu_{\mu}\to\nu_{e}$
and
$\nu_{\mu}\to\nu_{\tau}$
appearance
experiments.

Finally,
indications in favor of
$ \bar\nu_\mu \to \bar\nu_e $
oscillations
have been found in the LSND experiment
\cite{LSND}.
This is the only SBL experiment
which presently claims
an evidence in favor of neutrino oscillations.
The analysis of the data of this experiment,
taking into account
the negative results of other SBL experiments
(in particular,
the Bugey and BNL E776 experiments),
indicate a value of
$\delta{m}^{2}$
in the range
$
0.3
\lesssim
\delta{m}^{2}
\lesssim
2.2 \, \mathrm{eV}^2
$.

The three indications
in favor of neutrino oscillations
need three different scales of
$\delta{m}^{2}$,
which can be obtained with at least four massive neutrinos.
From the LEP measurements
of the invisible width of the $Z$-boson
(see Ref.\cite{PDG96})
we know that there are three light flavor neutrinos:
$\nu_e$,
$\nu_\mu$ and
$\nu_\tau$.
These are called flavor neutrinos
because they take part in weak interactions
and each of them couples to the corresponding
charged lepton through the charged-current
weak interaction.
However,
if the neutrino Lagrangian has a mass term,
in general the left-handed flavor neutrino
fields
$\nu_{{\alpha}L}$
are superpositions
of
the left-handed components
$\nu_{kL}$
of the fields of neutrinos with definite mass
($k=1,2,3,\ldots,n$):
\begin{equation}
\nu_{{\alpha}L}
=
\sum_{k=1}^{n}
U_{{\alpha}k}
\,
\nu_{kL}
\,.
\label{102}
\end{equation}
Here $U$
is a unitary mixing matrix.
The number $n$ of massive neutrinos can be three or more,
without any experimental upper limit.
If $n>3$ there are
$n-3$ sterile neutrino fields,
i.e.,
fields of
neutrinos which do not take part in weak interactions.
Hence,
in Eq.(\ref{102}) we have
$\alpha=e,\mu,\tau,s_{1},s_{2},\ldots,s_{n-3}$,
with
$n-3$ sterile neutrino fields
$\nu_{s_{1}}$,
$\nu_{s_{2}}$,
\ldots,
$\nu_{s_{n-3}}$.

In the following we will consider mixing schemes
in which
only the largest mass square difference
$ \Delta{m}^{2}_{n1} \equiv m^2_n - m^2_1 $
is relevant for SBL oscillations
\cite{BBGK,BGKP,BGG96}.
These schemes are based on mass spectra with
two groups of massive neutrinos
with close masses,
$ \nu_1, \ldots , \nu_{r-1} $
and
$ \nu_{r}, \ldots , \nu_n $,
separated by a mass difference in the eV range:
$
m_1 < \ldots < m_{r-1}
\ll
m_r < \ldots < m_n
$.
The general expression for the transition probability
of relativistic neutrinos
(see Refs.\cite{BP87-CWKim,Mohapatra-Pal})
\begin{equation}
P_{\nu_\alpha\to\nu_\beta}
=
\left|
\sum_{k=1}^{n}
U_{{\beta}k}
\,
U_{{\alpha}k}^{*}
\,
e^{
- i
\,
\frac{ \Delta{m}_{k1} L }{ 2 E }
}
\right|^2
\label{151}
\end{equation}
can be written as
\begin{equation}
P_{\nu_\alpha\to\nu_\beta}
=
\left|
\sum_{k=1}^{r-1}
U_{{\beta}k}
\,
U_{{\alpha}k}^{*}
\,
e^{
- i
\,
\frac{ \Delta{m}^{2}_{k1} L }{ 2 E }
}
+
e^{
- i
\,
\frac{ \Delta{m}^{2}_{n1} L }{ 2 E }
}
\sum_{k=r}^{n}
U_{{\beta}k}
\,
U_{{\alpha}k}^{*}
\,
e^{
i
\,
\frac{ \Delta{m}^{2}_{nk} L }{ 2 E }
}
\right|^2
\,,
\label{152}
\end{equation}
where
$L$ is the
distance between
the neutrino source and detector,
$E$ is the neutrino energy
and
$ \Delta{m}^{2}_{kj} \equiv m_k^2 - m_j^2 $.
In SBL experiments we have
\begin{equation}
\frac{ \Delta{m}^{2}_{n1} L }{ 2 E }
\gtrsim
1
\,,
\quad
\frac{ \Delta{m}^{2}_{k1} L }{ 2 E }
\ll
1
\quad \mbox{for} \quad
k < r
\quad \mbox{and} \quad
\frac{ \Delta{m}^{2}_{nk} L }{ 2 E }
\ll
1
\quad \mbox{for} \quad
k \geq r
\,,
\label{107}
\end{equation}
From Eqs.(\ref{153}) and (\ref{107}),
for the transition ($\beta\neq\alpha$)
and survival ($\beta=\alpha$) probabilities
of neutrinos (anti-neutrinos)
in SBL experiments
we have the following expressions
\cite{BGKP}:
\begin{equation}
P^{(\mathrm{SBL})}_{\stackrel{\makebox[0pt][l]
{$\hskip-3pt\scriptscriptstyle(-)$}}{\nu_{\alpha}}
\to\stackrel{\makebox[0pt][l]
{$\hskip-3pt\scriptscriptstyle(-)$}}{\nu_{\beta}}}
=
\frac{1}{2}
\,
A_{\alpha;\beta}
\left( 1 - \cos \frac{ \Delta{m}^{2} L }{ 2 E } \right)
,
\quad
P^{(\mathrm{SBL})}_{\stackrel{\makebox[0pt][l]
{$\hskip-3pt\scriptscriptstyle(-)$}}{\nu_{\alpha}}
\to\stackrel{\makebox[0pt][l]
{$\hskip-3pt\scriptscriptstyle(-)$}}{\nu_{\alpha}}}
=
1 - \frac{1}{2}
\,
B_{\alpha;\alpha}
\left(1 - \cos \frac{ \Delta{m}^{2} L }{ 2 E } \right)
,
\label{108}
\end{equation}
with
$ \Delta{m}^{2} \equiv \Delta{m}^{2}_{n1} $
and
the oscillation amplitudes
\begin{eqnarray}
&&
A_{\alpha;\beta}
=
4 \left| \sum_{k=r}^{n} U_{{\beta}k} \, U_{{\alpha}k}^{*} \right|^2
=
4 \left| \sum_{k=1}^{r-1} U_{{\beta}k} \, U_{{\alpha}k}^{*} \right|^2
,
\label{109}
\\
&&
B_{\alpha;\alpha}
=
4
\left( \sum_{k=r}^{n} |U_{{\alpha}k}|^2 \right)
\left( 1 - \sum_{k=r}^{n} |U_{{\alpha}k}|^2 \right)
=
4
\left( \sum_{k=1}^{r-1} |U_{{\alpha}k}|^2 \right)
\left( 1 - \sum_{k=1}^{r-1} |U_{{\alpha}k}|^2 \right)
.
\label{111}
\end{eqnarray}
The equalities
of the two expressions
in Eq.(\ref{109})
and
in Eq.(\ref{111})
are due to
the unitarity of the mixing matrix.
The formulas (\ref{108})
have the same form of the standard expressions
for the oscillation probabilities in the case of
two neutrinos
(see Refs.\cite{BP87-CWKim,Mohapatra-Pal}).
This fact is very important,
because the data of all the SBL experiments
have been analyzed by the experimental groups
under the assumption
of two-generation mixing,
obtaining constraints on the possible values of
the mixing parameters
$\Delta{m}^{2}$
and
$\sin^{2}2\theta$
($\theta$ is the mixing angle).
Hence,
we can use the results of the analyses of
the neutrino oscillation data
made by the experimental groups
in order to constraint
the possible values of
the oscillation amplitudes
$A_{\alpha;\beta}$
and
$B_{\alpha;\alpha}$.

First,
we consider the scheme with three neutrinos
and the mass hierarchy
(see also Ref.\cite{three})
\begin{equation}
\underbrace{
\overbrace{m_1 \ll m_2}^{\mathrm{solar}}
\ll
m_3
}_{\mathrm{LSND}}
\,.
\label{103}
\end{equation}
This scheme
(as all the schemes with three neutrinos)
provides only two independent
mass-squared differences,
$\Delta{m}^{2}_{21}$
and
$\Delta{m}^{2}_{31}$,
which
we choose to be relevant for the solution of the
solar neutrino problem
and
for neutrino oscillations in the LSND experiment.

Let us emphasize that the mass spectrum (\ref{103})
with three neutrinos and a mass hierarchy
is the simplest and most natural one,
being analogous to the mass spectra of
charged leptons, up and down quarks.
Moreover,
a scheme with three neutrinos and a mass hierarchy
is predicted by the see-saw mechanism
for the generation of neutrino masses
\cite{see-saw},
which can explain the smallness of the neutrino masses
with respect to the masses of the corresponding
charged leptons.

In the case of scheme (\ref{103})
we have
$n=r=3$
and
from Eqs.(\ref{109}) and (\ref{111})
it follows that
\begin{equation}
A_{\alpha;\beta}
=
4
\,
|U_{{\alpha}3}|^2
\,
|U_{{\beta}3}|^2
\,,
\qquad
B_{\alpha;\alpha}
=
4
\,
|U_{{\alpha}3}|^2
\left(
1
-
|U_{{\alpha}3}|^2
\right)
\,.
\label{AABB3}
\end{equation}
Hence,
neutrino oscillations in
SBL experiments
depend on three parameters:
$\Delta{m}^{2}\equiv\Delta{m}^{2}_{31}$,
$|U_{e3}|^2$ and
$|U_{\mu3}|^2$
(the unitarity of $U$ implies that
$
|U_{\tau3}|^2
=
1
-
|U_{e3}|^2
-
|U_{\mu3}|^2
$).

\begin{figure}[t]
\begin{minipage}[t]{0.49\linewidth}
\begin{center}
\mbox{\epsfig{file=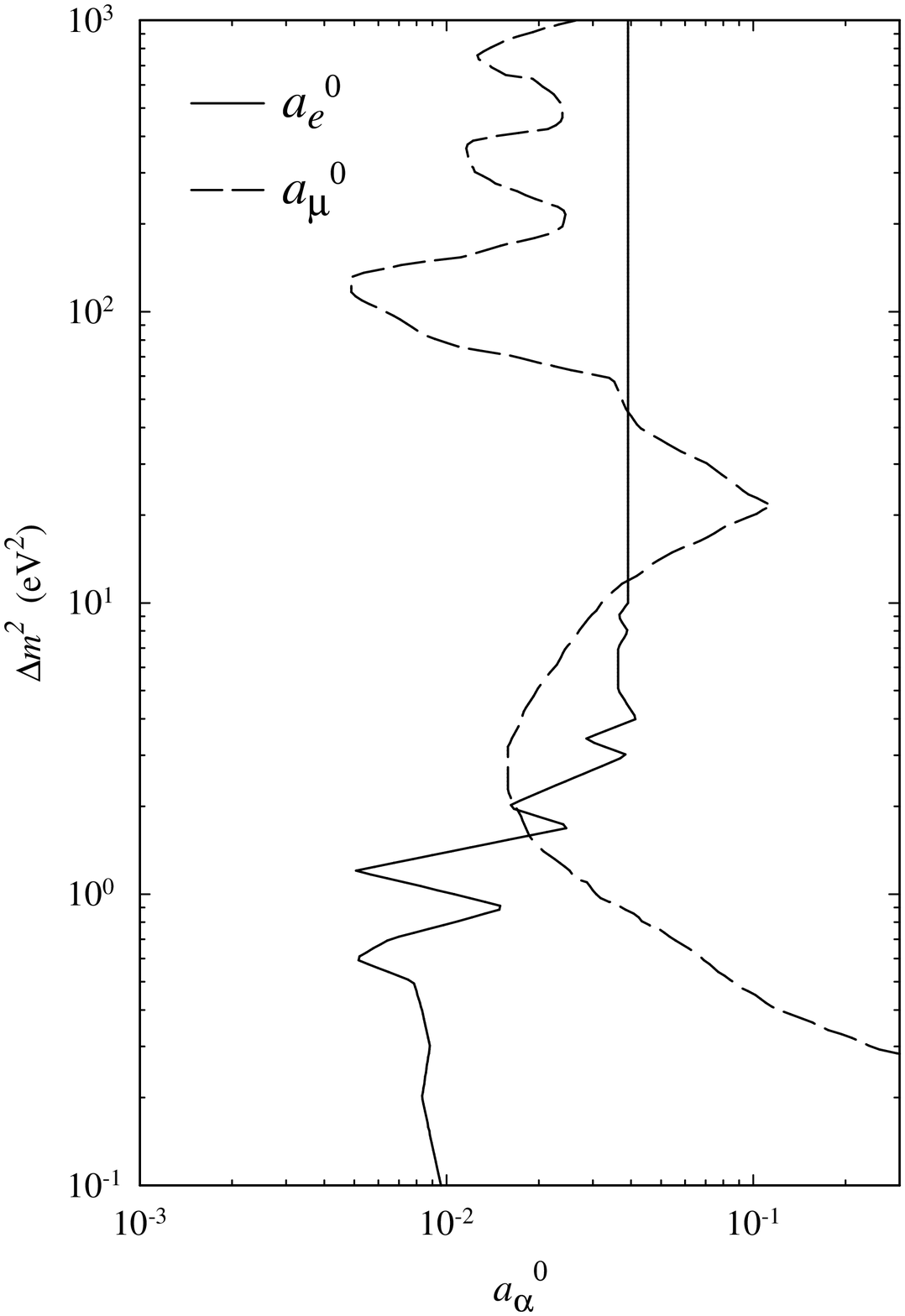,width=0.9\linewidth}}
\\
Figure \ref{fig1}
\end{center}
\end{minipage}
\null
\vspace{-0.3cm}
\null
\refstepcounter{figure}
\label{fig1}
\hfill
\begin{minipage}[t]{0.49\linewidth}
\begin{center}
\mbox{\epsfig{file=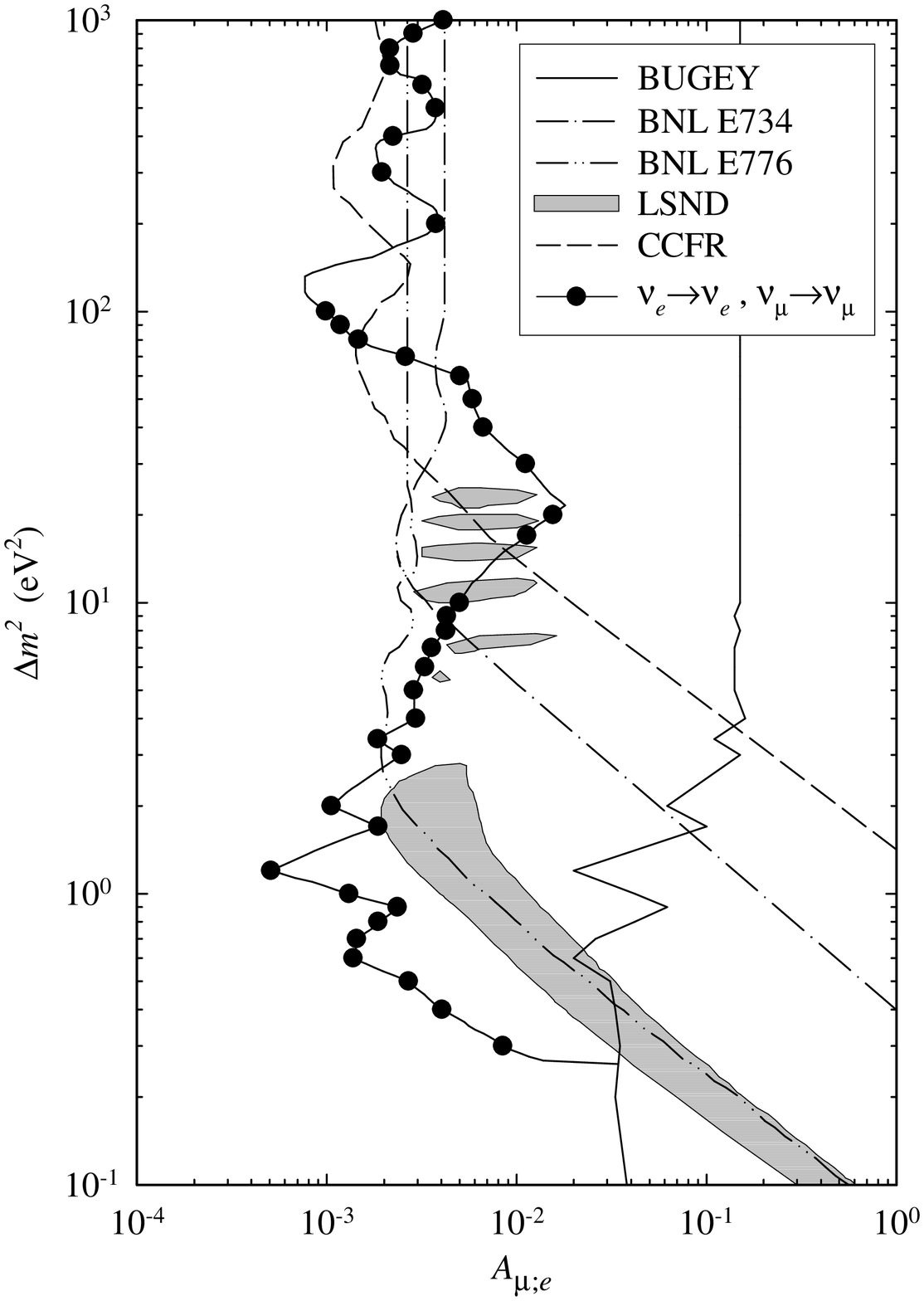,width=0.9\linewidth}}
\\
Figure \ref{fig2}
\end{center}
\end{minipage}
\null
\vspace{-0.3cm}
\null
\refstepcounter{figure}
\label{fig2}                 
\end{figure}

From the exclusion plots obtained in
reactor $\bar\nu_e$
and accelerator
$
\stackrel{\makebox[0pt][l]
{$\hskip-3pt\scriptscriptstyle(-)$}}{\nu_{\mu}}
$
disappearance experiments
it follows that,
at any fixed value of $\Delta{m}^2$,
the oscillation amplitudes
$B_{e;e}$
and
$B_{\mu;\mu}$
are bounded by the upper values
$B_{e;e}^{0}$
and
$B_{\mu;\mu}^{0}$,
respectively.
The values of
$B_{e;e}^{0}$
and
$B_{\mu;\mu}^{0}$
given by the exclusion plots obtained in the
Bugey \cite{Bugey95}
$\bar\nu_e$
disappearance experiment
and in the
CDHS and CCFR
\cite{CDHS84-CCFR84}
$
\stackrel{\makebox[0pt][l]
{$\hskip-3pt\scriptscriptstyle(-)$}}{\nu_{\mu}}
$
disappearance experiments
are small for any value of
$\Delta{m}^2$
in the wide interval
$
0.3
\lesssim
\Delta{m}^2
\lesssim
10^{3} \, \mathrm{eV}^2
$.
From Eq.(\ref{AABB3})
one can see that small upper bounds for
$B_{e;e}$
and
$B_{\mu;\mu}$
imply that the parameters
$|U_{e3}|^2$
and
$|U_{\mu3}|^2$
can be small or large (i.e., close to one):
\begin{equation}
|U_{{\alpha}3}|^2
\leq
a^0_\alpha
\qquad \mbox{or} \qquad
|U_{{\alpha}3}|^2
\geq
1 - a^0_\alpha
\qquad
(\alpha=e,\mu)
\,,
\label{uu}
\end{equation}
with
\begin{equation}
a^0_\alpha
=
\frac{1}{2}
\left(
1
-
\sqrt{ 1 - B_{\alpha;\alpha}^{0} }
\right)
\,.
\label{a0}
\end{equation}
Both
$a^{0}_{e}$
and
$a^{0}_{\mu}$
are small
($ a^{0}_e \lesssim 4 \times 10^{-2} $
and
$ a^{0}_\mu \lesssim 2 \times 10^{-1} $)
for any value of
$\Delta{m}^{2}$
in the range
$
0.3
\lesssim
\Delta{m}^2
\lesssim
10^{3} \, \mathrm{eV}^2
$
(see Fig.\ref{fig1}).

Since large values of both
$|U_{e3}|^2$
and
$|U_{\mu3}|^2$
are excluded by the unitarity of the mixing matrix
($|U_{e3}|^2+|U_{\mu3}|^2\leq1$),
at any fixed value of $\Delta{m}^2$
there are three regions in the
$|U_{e3}|^2$--$|U_{\mu3}|^2$
plane
which are allowed by
the exclusion plots of SBL disappearance experiments:
\emph{Region} I,
with
$ |U_{e3}|^2 \leq a^{0}_{e} $
and
$ |U_{\mu3}|^2 \leq a^{0}_{\mu} $;
\emph{Region} II,
with
$ |U_{e3}|^2 \leq a^{0}_{e} $
and
$ |U_{\mu3}|^2 \geq 1 - a^{0}_{\mu} $;
\emph{Region} III,
with
$ |U_{e3}|^2 \geq 1 - a^{0}_{e} $
and
$ |U_{\mu3}|^2 \leq a^{0}_{\mu} $.

In Region III
$ |U_{e3}|^2 $
is large and $\nu_e$
has a large mixing with
$\nu_3$
and a small mixing with $\nu_1$ and $\nu_2$.
Since the squared-mass difference
$\Delta{m}^{2}_{21}$
is assumed to be responsible
for the oscillations of solar neutrinos,
a small mixing of the electron neutrino
with $\nu_1$ and $\nu_2$
implies that the oscillations
of solar $\nu_e$'s
are suppressed and the solar neutrino problem cannot
be solved by neutrino oscillations.
Indeed,
the survival probability of solar $\nu_e$'s,
$P_{\nu_e\to\nu_e}^{\mathrm{sun}}$,
is bounded by
$
P_{\nu_e\to\nu_e}^{\mathrm{sun}}
\geq
|U_{e3}|^4
$
(see Ref.\cite{BBGK}).
If 
$ |U_{e3}|^2 \geq 1 - a^0_e $,
we have
$
P_{\nu_e\to\nu_e}^{\mathrm{sun}}
\geq
0.92
$
at all neutrino energies,
which is a bound that is not compatible
with the solar neutrino data.
Hence,
Region III is excluded by solar neutrinos.

The Region I
is disfavored by the results of the LSND experiment.
Indeed,
in Region I we have
\begin{equation}
A_{\mu;e} \leq 4 \, a^0_e \, a^0_\mu
\,.
\label{amuel}
\end{equation}
This inequality implies that
$
\stackrel{\makebox[0pt][l]
{$\hskip-3pt\scriptscriptstyle(-)$}}{\nu_{\mu}}
\leftrightarrows
\stackrel{\makebox[0pt][l]
{$\hskip-3pt\scriptscriptstyle(-)$}}{\nu_{e}}
$
transitions in SBL experiments are strongly suppressed.
The upper bound obtained
with the inequality (\ref{amuel})
from the 90\% CL exclusion plots of the
Bugey \cite{Bugey95}
$\bar\nu_e$
disappearance experiment
and of the
CDHS and CCFR \cite{CDHS84-CCFR84}
$
\stackrel{\makebox[0pt][l]
{$\hskip-3pt\scriptscriptstyle(-)$}}{\nu_{\mu}}
$
disappearance experiments
is represented in Fig.\ref{fig2}
by the curve passing trough the circles.
The shadowed regions in Fig.\ref{fig2}
are allowed at 90\% CL by the results of the LSND experiment.
Also shown
are the 90\% CL exclusion curves found in the
BNL E734,
BNL E776
and
CCFR
\cite{BNLE734-BNLE776-CCFR96}
$
\stackrel{\makebox[0pt][l]
{$\hskip-3pt\scriptscriptstyle(-)$}}{\nu_{\mu}}
\to
\stackrel{\makebox[0pt][l]
{$\hskip-3pt\scriptscriptstyle(-)$}}{\nu_{e}}
$
appearance experiments
and in the Bugey experiment.
One can see from Fig.\ref{fig2} that
the bounds obtained from the direct experiments on the search for
$
\stackrel{\makebox[0pt][l]
{$\hskip-3pt\scriptscriptstyle(-)$}}{\nu_{\mu}}
\to
\stackrel{\makebox[0pt][l]
{$\hskip-3pt\scriptscriptstyle(-)$}}{\nu_{e}}
$
oscillations and the bound
(\ref{amuel})
obtained in Region I
are not compatible
with the allowed regions of the LSND experiment
\cite{BBGK}.
Therefore,
we come to the conclusion that the Region I
is not favored by the existing experimental data.
This is an important result,
because
the Region I
is the only one in which it is possible to have
a hierarchy of
the elements of the neutrino mixing matrix
analogous to the one of the
quark mixing matrix.

Having excluded the Regions I and III
of the scheme (\ref{103}),
we are left only with the Region II,
where $\nu_\mu$ has a large mixing
with $\nu_3$,
i.e.,
$\nu_\mu$
(not $\nu_\tau$)
is the ``heaviest'' neutrino.

The scheme (\ref{103})
does not allow to explain
the atmospheric neutrino anomaly
with neutrino oscillations.
However,
it must be noticed that only
the fit of the Kamiokande multi-GeV data sample
requires a definite value of
$\delta{m}^{2}$
of the order of
$ 10^{-2} \, \mathrm{eV}^2 $,
whereas the Kamiokande sub-GeV data
and the IMB and Soudan data
can be fitted also with higher values of
$\delta{m}^{2}$.
Hence,
in the three neutrino scheme (\ref{103}) a
$ \Delta{m}^{2}_{31} \simeq 0.3 \, \mathrm{eV}^2 $,
which is the lowest value allowed by the results
of the Bugey and LSND experiments,
could be responsible also for the oscillations
of atmospheric neutrinos \cite{CF97-MR97}.
Indeed,
the CDHS exclusion curve implies that
$ B_{\mu;\mu}^{0} \simeq 0.7 $
at
$ \Delta{m}^{2}_{31} \simeq 0.3 \, \mathrm{eV}^2 $.
For the sub-GeV events we have
$
P^{\mathrm{atm}}_{\nu_{\mu}\to\nu_{\mu}}
=
1 - \frac{1}{2}
\,
B_{\mu;\mu}
$.
Therefore, we get the lower bound
$
P^{\mathrm{atm}}_{\nu_{\mu}\to\nu_{\mu}}
\gtrsim
0.65
$.
One can show \cite{BGG96}
that the double-ratio
$
R = (\mu/e)_{\mathrm{data}}/(\mu/e)_{\mathrm{MC}}
$
of sub-GeV muon and electron events
in the Kamiokande detector
($(\mu/e)_{\mathrm{MC}}$
is the Monte-Carlo
calculated ratio of muon and electron events
without neutrino oscillations)
satisfies the inequality
$
R
\geq
P^{\mathrm{atm}}_{\nu_{\mu}\to\nu_{\mu}}
$.
Therefore,
for
$ \Delta{m}^{2}_{31} \simeq 0.3 \, \mathrm{eV}^2 $
we have
$ R \gtrsim 0.65 $,
which is compatible with the experimental value
$
R^{\mathrm{exp}}
=
0.60^{+0.06}_{-0.05} \pm 0.05
$.
At higher values of
$ \Delta{m}^{2}_{31} $
the CDHS exclusion curve gives lower values for
$ B_{\mu;\mu}^{0} $,
leading to an incompatibility with the Kamiokande sub-GeV data.
For example,
at
$ \Delta{m}^{2}_{31} \simeq 0.4 \, \mathrm{eV}^2 $
from the CDHS exclusion curve we have
$ B_{\mu;\mu}^{0} \simeq 0.4 $,
which yields the lower bound
$ R \gtrsim 0.80 $
that is incompatible with the experimental value.

Another scheme with three neutrinos and a mass hierarchy
in which
$ \Delta{m}^{2}_{21} \simeq 10^{-2} \, \mathrm{eV}^2 $
is responsible for the oscillations
of solar and atmospheric neutrinos
has been proposed in Ref.\cite{AP97}.
This scheme corresponds to Region I,
which is marginally allowed for
$ \Delta{m}^{2}_{31} \simeq 1.7 \, \mathrm{eV}^2 $,
as can be seen from Fig.\ref{fig2}.
However,
the fact that in this scheme
the survival probability of solar $\nu_e$'s
does not depend on the neutrino energy
is at odds with the present solar neutrino data
\cite{KP96}.

Let us now consider the schemes with four neutrinos,
which provide three independent
mass-squared differences
and allow to accommodate
in a natural way
all the three experimental indications
in favor of neutrino oscillations
(see also Ref.\cite{four}).
We consider first the scheme with four neutrinos and
the mass hierarchy
\begin{equation}
\underbrace{
\overbrace{
\overbrace{m_1 \ll m_2}^{\mathrm{solar}}
\ll m_3
}^{\mathrm{atm}}
\ll m_4
}_{\mathrm{LSND}}
\,.
\label{104}
\end{equation}
The three independent
mass-squared differences,
$\Delta{m}^{2}_{21}$,
$\Delta{m}^{2}_{31}$ and
$\Delta{m}^{2}_{41}$,
are taken to be relevant for the oscillations
of solar, atmospheric and LSND neutrinos,
respectively.
In the case of scheme (\ref{104})
we have
$n=r=4$
and Eqs.(\ref{109}) and (\ref{111})
imply that
the oscillation amplitudes are given by
\begin{equation}
A_{\alpha;\beta}
=
4
\,
|U_{{\alpha}4}|^2
\,
|U_{{\beta}4}|^2
\,,
\qquad
B_{\alpha;\alpha}
=
4
\,
|U_{{\alpha}4}|^2
\left(
1
-
|U_{{\alpha}4}|^2
\right)
\,.
\label{AABB4}
\end{equation}
In this case,
neutrino oscillations in
SBL experiments
depend on four parameters:
$\Delta{m}^{2}\equiv\Delta{m}^{2}_{31}$,
$|U_{e3}|^2$,
$|U_{\mu3}|^2$ and
$|U_{\tau3}|^2$.
From the similarity of the amplitudes
(\ref{AABB4})
with the corresponding ones
given in Eq.(\ref{AABB3}),
it is clear that,
replacing
$|U_{{\alpha}3}|^2$
with
$|U_{{\alpha}4}|^2$,
we can apply
to the scheme (\ref{104})
the same analysis presented for the scheme (\ref{103})
with three neutrinos and a mass hierarchy.
At any fixed value of $\Delta{m}^2$
we have three regions in the
$|U_{e4}|^2$--$|U_{\mu4}|^2$
plane
which are allowed by the exclusion plots
of SBL disappearance experiments:
\emph{Region} I,
with
$ |U_{e4}|^2 \leq a^{0}_{e} $
and
$ |U_{\mu4}|^2 \leq a^{0}_{\mu} $;
\emph{Region} II,
with
$ |U_{e4}|^2 \leq a^{0}_{e} $
and
$ |U_{\mu4}|^2 \geq 1 - a^{0}_{\mu} $;
\emph{Region} III,
with
$ |U_{e4}|^2 \geq 1 - a^{0}_{e} $
and
$ |U_{\mu4}|^2 \leq a^{0}_{\mu} $.
The Regions III and I are excluded,
respectively,
by the solar neutrino problem
and by the results of the LSND experiment,
for the same reasons discussed in the case of the scheme (\ref{103}).
Furthermore,
the purpose of considering the scheme (\ref{104})
is to have the possibility to explain the atmospheric neutrino anomaly,
but this is not possible if the
neutrino mixing parameters lie in Region II.
Indeed,
in Region II
$ |U_{\mu4}|^2 $
is large
and the muon neutrino has a large mixing with the heaviest
massive neutrino, $\nu_4$,
and a small mixing with the light neutrinos
$\nu_1$,
$\nu_2$ and
$\nu_3$.
Since the atmospheric neutrino oscillations
are assumed to be due to the phase generated by
$\Delta{m}^{2}_{31}$,
a relatively large mixing of $\nu_\mu$
with the three light neutrinos
is necessary in order to explain
the observed deficit of atmospheric muon neutrinos.
In Ref.\cite{BGG96}
it has been shown quantitatively
that the small mixing
of $\nu_\mu$
with
$\nu_1$,
$\nu_2$ and
$\nu_3$
in Region II
is incompatible with the atmospheric neutrino data.

Hence,
in the framework of scheme (\ref{104})
all the regions of the mixing parameters
are incompatible with the results of
neutrino oscillation experiments
and we conclude that this scheme
is not favored by the experimental data.
It is possible to show that,
for the same reasons,
all possible schemes with four neutrinos
and a mass spectrum in which
three masses are clustered and one mass is separated from the
others by a gap of about 1 eV
(needed for the explanation of the LSND data)
are not compatible with the results
of all neutrino oscillation experiments.
Therefore,
there are only two possible schemes
with four neutrinos
which are compatible with the results
of all the neutrino oscillation experiments:
\begin{equation}
\mbox{(A)}
\qquad
\underbrace{
\overbrace{m_1 < m_2}^{\mathrm{atm}}
\ll
\overbrace{m_3 < m_4}^{\mathrm{solar}}
}_{\mathrm{LSND}}
\qquad \mbox{and} \qquad
\mbox{(B)}
\qquad
\underbrace{
\overbrace{m_1 < m_2}^{\mathrm{solar}}
\ll
\overbrace{m_3 < m_4}^{\mathrm{atm}}
}_{\mathrm{LSND}}
\,.
\label{105}
\end{equation}
In these two schemes
the four neutrino masses
are divided in two pairs of close masses
separated by a gap of about 1 eV.
In scheme A,
$\Delta{m}^{2}_{21}$
is relevant
for the explanation of the atmospheric neutrino anomaly
and
$\Delta{m}^{2}_{43}$
is relevant
for the suppression of solar $\nu_e$'s.
In scheme B,
the roles of
$\Delta{m}^{2}_{21}$
and
$\Delta{m}^{2}_{43}$
are reversed.

From Eq.(\ref{111}),
the oscillation amplitudes
$B_{\alpha;\alpha}$
in the schemes (\ref{105}),
with $n=4$ and $r=3$,
are given by
\begin{equation}
B_{\alpha;\alpha}
=
4 \, c_{\alpha} \left( 1 - c_{\alpha} \right)
\,,
\label{153}
\end{equation}
with the following definitions
of the parameters $c_{\alpha}$
in the two schemes A and B:
\begin{equation}
\mbox{(A)}
\qquad
c_{\alpha}
\equiv
\sum_{k=1,2}
|U_{{\alpha}k}|^2
\qquad \mbox{and} \qquad
\mbox{(B)}
\qquad
c_{\alpha}
\equiv
\sum_{k=3,4}
|U_{{\alpha}k}|^2
\,.
\label{154}
\end{equation}
The expression (\ref{153}) for
$B_{\alpha;\alpha}$
has the same form as the one in Eq.(\ref{AABB4}),
with
$|U_{{\alpha}4}|^2$
replaced by $c_{\alpha}$.
Therefore,
we can apply the same analysis to the results
of SBL disappearance experiments
as that presented for the case of scheme (\ref{104})
and we obtain four allowed regions
in the
$c_e$--$c_\mu$
plane
(now the region with large
$c_e$ and $c_\mu$
is not excluded by
the unitarity of the mixing matrix,
which gives the constraint
$ c_e + c_\mu \leq 2 $):
\emph{Region} I,
with
$ c_e \leq a^{0}_{e} $
and
$ c_\mu \leq a^{0}_{\mu} $;
\emph{Region} II,
with
$ c_e \leq a^{0}_{e} $
and
$ c_\mu \geq 1 - a^{0}_{\mu} $;
\emph{Region} III,
with
$ c_e \geq 1 - a^{0}_{e} $
and
$ c_\mu \leq a^{0}_{\mu} $;
\emph{Region} IV,
with
$ c_e \geq 1 - a^{0}_{e} $
and
$ c_\mu \geq 1 - a^{0}_{\mu} $.
However,
following the same reasoning
as in the case of scheme (\ref{104}),
one can see that
the Regions III and IV
are excluded by the solar neutrino data
and the Regions I and III
are excluded by the results of the atmospheric neutrino experiments
\cite{BGG96}.
Hence,
only the Region II
is allowed by the results of all experiments.

If the neutrino mixing parameters lie in Region II,
in the scheme A (B) the electron (muon)
neutrino is ``heavy'',
because it has a large mixing with $\nu_3$ and $\nu_4$,
and the muon (electron) neutrino is light.
Thus,
the schemes A and B give different predictions
for the effective Majorana mass
$
\langle{m}\rangle
=
\sum_{k} U_{ek}^2 m_k
$
in
neutrinoless double-beta decay
experiments:
since
$ m_3 \simeq m_4 \gg m_1 \simeq m_2 $,
we have
\begin{equation}
(\mbox{A})
\quad
|\langle{m}\rangle|
\leq
(1-c_{e}) m_4
\simeq
m_4
\,,
\qquad
(\mbox{B})
\quad
|\langle{m}\rangle|
\leq
c_e m_4
\leq
a^{0}_{e} m_4
\ll m_4
\,.
\label{155}
\end{equation}
Thus, if scheme A is realized in nature
the experiments on the search for neutrinoless double-beta
decay can reveal
the effects of the heavy
neutrino masses $ m_3 \simeq m_4 $.
Furthermore,
the smallness of $c_e$
in both schemes A and B
implies that the electron neutrino has a
small mixing with the neutrinos whose mass-squared difference is
responsible for the oscillations of atmospheric and LBL neutrinos
(i.e.,
$\nu_1$, $\nu_2$ in scheme A and $\nu_3$, $\nu_4$ in scheme
B). Hence, the probability of transitions of atmospheric and
LBL electron neutrinos into other states is suppressed
\cite{BGG97}.

In conclusion,
the analysis presented here
shows that,
if the experimental indications
in favor of neutrino oscillations
are confirmed,
the mixing of leptons
is very different from the mixing of quarks.


\begin{thebibliography}{99}

\bibitem{Pontecorvo57}
B. Pontecorvo,
J. Exptl. Theoret. Phys. \textbf{33}, 549 (1957)
[Sov. Phys. JETP \textbf{6}, 429 (1958)];
J. Exptl. Theoret. Phys. \textbf{34}, 247 (1958)
[Sov. Phys. JETP \textbf{7}, 172 (1958)].

\bibitem{BP87-CWKim}
S.M. Bilenky and S.T. Petcov,
Rev. Mod. Phys. \textbf{59}, 671 (1987);
C.W. Kim and A. Pevsner,
\emph{Neutrinos in Physics and Astrophysics},
Contemporary Concepts in Physics, Vol. 8
(Harwood Academic Press, Chur, Switzerland, 1993).

\bibitem{Mohapatra-Pal}
R.N. Mohapatra and P.B. Pal,
\emph{Massive Neutrinos in Physics and Astrophysics},
World Scientific Lecture Notes in Physics, Vol. 41
(Singapore, 1991).

\bibitem{Bugey95}
B. Achkar \emph{et al.},
Nucl. Phys. B \textbf{434}, 503 (1995).

\bibitem{CDHS84-CCFR84}
F. Dydak \emph{et al.},
Phys. Lett. B \textbf{134}, 281 (1984);
I.E. Stockdale \emph{et al.},
Phys. Rev. Lett. \textbf{52}, 1384 (1984).

\bibitem{BNLE734-BNLE776-CCFR96}
L.A. Ahrens \emph{et al.},
Phys. Rev. D \textbf{36}, 702 (1987);
L. Borodovsky \emph{et al.},
Phys. Rev. Lett. \textbf{68}, 274 (1992);
A. Romosan \emph{et al.},
Phys. Rev. Lett. \textbf{78}, 2912 (1997).

\bibitem{solarexp}
B.T. Cleveland \emph{et al.},
Nucl. Phys. B (Proc. Suppl.) \textbf{38}, 47 (1995);
K.S. Hirata \emph{et al.},
Phys. Rev. D \textbf{44}, 2241 (1991);
GALLEX Coll.,
Phys. Lett. B \textbf{388}, 384 (1996);
J.N. Abdurashitov \emph{et al.},
Phys. Rev. Lett. \textbf{77}, 4708 (1996);
Y. Takeuchi,
Talk presented at
32$^{\mathrm{nd}}$
\emph{Rencontres de Moriond: Electroweak Interactions and Unified Theories},
Les Arcs, France, March 1997.

\bibitem{SSM}
J.N. Bahcall and M.H. Pinsonneault,
Rev. Mod. Phys. \textbf{67}, 781 (1995);
S. Turck-Chi\`eze \emph{et al.},
Phys. Rep. \textbf{230}, 57 (1993);
V. Castellani \emph{et al.},
Phys. Rep. \textbf{281}, 309 (1997);
A. Dar and G. Shaviv,
Nucl. Phys. B (Proc. Suppl.) \textbf{48}, 335 (1996).

\bibitem{MSW}
S.P. Mikheyev and A.Yu. Smirnov,
Yad. Fiz. \textbf{42}, 1441 (1985)
[Sov. J. Nucl. Phys. \textbf{42}, 913 (1985)];
L. Wolfenstein,
Phys. Rev. D \textbf{17}, 2369 (1978).

\bibitem{Kamiokande-IMB-Soudan}
Y. Fukuda \emph{et al.},
Phys. Lett. B \textbf{335}, 237 (1994);
R. Becker-Szendy \emph{et al.},
Nucl. Phys. B (Proc. Suppl.) \textbf{38}, 331 (1995);
W.W.M. Allison \emph{et al.},
Phys. Lett. B \textbf{391}, 491 (1997).

\bibitem{Frejus-NUSEX}
K. Daum \emph{et al.},
Z. Phys. C \textbf{66}, 417 (1995);
M. Aglietta \emph{et al.},
Europhys. Lett. \textbf{15}, 559 (1991).

\bibitem{up-mu}
M. Mori \emph{et al.},
Phys. Lett. B \textbf{270}, 89 (1991).;
R. Becker-Szendy \emph{et al.},
Phys. Rev. Lett. \textbf{69}, 1010 (1992);
M.M. Boliev \emph{et al.},
Proc. of the
$3^{\mathrm{th}}$ \emph{International Workshop
on Neutrino Telescopes},
Venezia, March 1991;
MACRO Coll.,
Phys. Lett. B \textbf{357}, 481 (1995).

\bibitem{K2K}
Y. Suzuki,
Talk presented at
\emph{Neutrino 96},
Helsinki, June 1996.

\bibitem{CHOOZ-PaloVerde}
R.I. Steinberg,
Proc. of the
\emph{$5^{\mathrm{th}}$ International Workshop
on Neutrino Telescopes},
Venezia, March 1993;
F. Boehm \emph{et al.},
\emph{The Palo Verde experiment},
1996.

\bibitem{MINOS-ICARUS}
MINOS Coll.,
NUMI-L-63, Feb. 1995;
ICARUS Coll.,
LNGS-94/99-I,
May 1994.

\bibitem{LSND}
C. Athanassopoulos \emph{et al.},
Phys. Rev. Lett. \textbf{77}, 3082 (1996).

\bibitem{PDG96}
R.M. Barnett \emph{et al.},
Phys. Rev. D \textbf{54}, 1 (1996).

\bibitem{BBGK}
S.M. Bilenky \emph{et al.},
Phys. Lett. B \textbf{356}, 273 (1995);
Phys. Rev. D \textbf{54}, 1881 (1996).

\bibitem{CF97-MR97}
C.Y. Cardall and G.M. Fuller,
Phys. Rev. D \textbf{53}, 4421 (1996);
C.Y. Cardall, G.M. Fuller and D.B. Cline,
hep-ph/9706426;
E. Ma and P. Roy,
hep-ph/9706309.

\bibitem{BGKP}
S.M. Bilenky \emph{et al.},
Phys. Rev. D \textbf{54}, 4432 (1996).

\bibitem{three}
A. De Rujula \emph{et al.},
Nucl. Phys. B \textbf{168}, 54 (1980);
V. Barger and K. Whisnant,
Phys. Lett. B \textbf{209}, 365 (1988);
S.M. Bilenky \emph{et al.},
\emph{ibid.} \textbf{276}, 223 (1992);
K.S. Babu \emph{et al.},
\emph{ibid.} \textbf{359}, 351 (1995).
H. Minakata,
\emph{ibid.} \textbf{356}, 61 (1995);
Phys. Rev. D \textbf{52}, 6630 (1995);
G.L. Fogli \emph{et al.},
\emph{ibid.} \textbf{52}, 5334 (1995);
S. Goswami \emph{et al.},
Int. J. Mod. Phys. A \textbf{12}, 781 (1997).

\bibitem{see-saw}
M. Gell-Mann, P. Ramond, and R. Slansky,
in \emph{Supergravity},
North Holland, Amsterdam, 1979, p.315;
S. Weinberg,
Phys. Rev. Lett. \textbf{43}, 1566 (1979).

\bibitem{BGG96}
S.M. Bilenky, C. Giunti and W. Grimus,
hep-ph/9607372.

\bibitem{AP97}
A. Acker and S. Pakvasa,
Phys. Lett. B \textbf{357}, 209 (1997).

\bibitem{KP96}
P.I. Krastev and S.T. Petcov,
Phys. Rev. D \textbf{53}, 1665 (1996).

\bibitem{four}
J.T. Peltoniemi and J.W.F. Valle,
Nucl. Phys. B \textbf{406}, 409 (1993);
D.O. Caldwell and R.N. Mohapatra,
Phys. Rev. D \textbf{48}, 3259 (1993);
\emph{ibid.} \textbf{50}, 3477 (1994);
Z. Berezhiani and R.N. Mohapatra,
\emph{ibid.} \textbf{52}, 6607 (1995);
E. Ma and P. Roy,
\emph{ibid.} \textbf{52}, R4780 (1995);
R. Foot and R.R. Volkas,
\emph{ibid.} \textbf{52}, 6595 (1995);
S. Goswami,
\emph{ibid.} \textbf{55}, 2931 (1997);
A.Yu. Smirnov and M. Tanimoto,
\emph{ibid.} \textbf{55}, 1665 (1997);
J.R. Primack \emph{et al.},
Phys. Rev. Lett. \textbf{74}, 2160 (1995);
E.J. Chun \emph{et al.},
Phys. Lett. B \textbf{357}, 608 (1995);
J.J. Gomez-Cadenas and M.C. Gonzalez-Garcia,
Z. Phys. C \textbf{71}, 443 (1996);
E. Ma,
Mod. Phys. Lett. A \textbf{11}, 1893 (1996).

\bibitem{BGG97}
S.M. Bilenky, C. Giunti and W. Grimus,
hep-ph/9705300.

\end{thebibliography}
\end{document}